# 108 ps coincidence time resolution through optimized scintillators, photodetectors, readout electronics, and DOI-based timing correction in orthogonally stacked detector configurations

**Arisa Sanzen[1], Yuya Onishi[1], Takahiro Moriya, Tomohide Omura and Ryosuke Ota[*]**

Central Research Laboratory, Hamamatsu Photonics K. K., 5000 Hirakuchi, Hamana-ku, Hamamatsu 434-8601, Japan
[1] These authors contributed equally to this work.

*Author to whom any correspondence should be addressed.

**E-mail:** ryosuke.ota@crl.hpk.co.jp

## Abstract

*Objective.* Existing commercial time-of-flight positron emission tomography (TOF-PET) systems yield a coincidence time resolution (CTR) of ~200 ps or less full width at half maximum (FWHM). Recently, there has been a challenge to achieve a CTR of 100 ps FWHM at the system level. However, current silicon photomultipliers (SiPMs) and 20-mm-thick scintillators in conventional single-ended readout scheme is difficult to achieve 100 ps CTR; the photon transport time spread (PTS) within the scintillator crystal is a major barrier. Differences in the interaction position result in variations in PTS on the order of several tens of ps, thereby degrading the CTR. A shorter scintillator can improve CTR; however, this can degrade detection efficiency. *Approach.* To overcome this trade-off between the CTR and detection efficiency, we previously proposed xDetector, an orthogonally stacked configuration along the longitudinal axis of scintillator crystals. We investigated the CTR potential of the xDetector by improving the scintillator, photodetector, and readout electronics, and by applying CTR correction based on a three-dimensional interaction within the scintillator. *Main results.* Based on error propagation, the CTR of the paired xDetector was calculated as $113.5 \pm 2.7$ ps FWHM. Furthermore, the CTR of the xDetector was measured at four positions along the longitudinal axis by manually sliding the xDetector, and the corrected achieved CTR was $108.6 \pm 1.9$ ps FWHM. Moreover, compared with the conventional single detector using a 20.0 mm scintillator, CTR improved by an average of 10.3%. *Significance.* The xDetector offers potential as a PET detector concept to achieve a CTR of 100 ps FWHM. Such timing performance is expected to improve TOF-PET image quality and quantitative accuracy, contributing to more reliable disease detection and diagnosis than current PET detectors.

## 1. Introduction

Positron emission tomography (PET) is a nuclear imaging technique that estimates annihilation positions by detecting back-to-back 511 keV gamma rays within the patient body using PET detectors arranged in a ring. PET is routinely employed for diagnosing various conditions, including diseases such as cancer, and neurodegenerative disorders such as Alzheimer's disease. The quality of PET images can be enhanced by incorporating time-of-flight (TOF) information from coincidence events (Schaart 2021).

The signal-to-noise ratio (SNR) gain between non-TOF and TOF PET is expressed as

$$SNR\ Gain = \frac{SNRTOF}{SNRnon{-}TOF} = \sqrt{\frac{2*D}{c*\Delta t}} \tag{1}$$

where $D$, $c$, and Δt represent the diameter of a patient, the speed of light, and coincidence time resolution (CTR), respectively. As shown in Equation (1), improving the CTR explicitly enhances the image SNR, thereby improving image quality. Thus, advancing CTR performance is necessary because it translates to a more accurate disease localization, shorter scan times, and/or reduced patient radiation dose.

Currently, commercially available TOF-PET systems achieve a CTR of ~200 ps FWHM (van Suluis *et al.* 2019, Zhang *et al.* 2024). Although these systems yield PET images with improved SNR compared to those of non-TOF PET systems, further improvements in CTR are necessary for enhancing diagnostic accuracy (Lecoq *et al.* 2020). Consequently, many studies are being conducted to achieve a 100 ps FWHM CTR (Pourashraf *et al.* 2025, Namallapudi *et al.* 2015, Schaart *et al.* 2010).

The analytical CTR can be expressed as (Gundacker *et al.* 2020)

$$CTRanalytic \propto \sqrt{\frac{\tau_d * (1.57 * \tau_r + 1.13 * \sigma_{SPTR+PTS})}{PDE * LTE * ILY@Energy}} \quad (2)$$

where $\tau_d$, $\tau_r$, $\sigma_{SPTR+PTS}$, $PDE$, $LTE$, and $ILY$ represent the effective decay time of the scintillator, scintillation rise time, convolution of the single photon time resolution (SPTR) of a silicon photomultiplier (SiPM) and photon transport spread (PTS) in the scintillator, photon detection efficiency of the SiPM, light transfer efficiency, and intrinsic light yield of the scintillator, respectively. As shown in Equation (2), CTR explicitly depends on the performance of the SiPM, scintillator, and detector geometry.

Assuming all other factors remain constant, a four-fold increase in PDE would be required to improve the CTR from 200 to 100 ps FWHM. However, current SiPMs already achieve PDEs exceeding 55%; this makes such an improvement impossible. Moreover, the contribution of SPTR improvement to the CTR can be limited when the PTS is dominant because SPTR is convolved with PTS. Although applying a metal mask to the SiPM has been reported to improve SPTR by ~30%, masking the periphery of the SiPM reduces the effective photosensitive area, resulting in ~30% decrease in PDE. This consequently degrades the CTR from 106 to 115 ps FWHM when using a 3 × 3 × 20 mm LYSO:Ce,Mg crystal (Gundacker *et al.* 2023). Thus, enhancing explicit parameters such as PDE is a more effective approach for improving the CTR. Parameters $\tau_d$, $\tau_r$, and $ILY$ are intrinsic material properties; $\tau_d$ and $ILY$ can be alternatively controlled by the amount of dopant. However, simultaneously achieving a fast $\tau_d$ and high ILY is a considerable challenge.

In current clinical PET systems, a 20-mm-thick scintillator is required to maintain a sufficient detection efficiency for 511 keV gamma rays (van Sluis *et al.* 2019; Hsu *et al.* 2017). However, a 20 mm thickness introduces ~75 ps of PTS, which poses a significant barrier for achieving a 100 ps FWHM (Gundacker *et al.* 2014; Cates *et al.* 2015). Thus, a straightforward strategy for reducing PTS is using thinner scintillators. Recent studies have reported achieving sub-100 ps FWHM using thinner scintillators in combination with improvements in photodetectors and readout electronics (Mariscal-Castilla *et al.* 2024; Gundacker *et al.* 2019). Although thinner scintillators reduce PTS and enhance $LTE$ (Cates and Levin 2018), they decrease detection efficiency. Consequently, employing thinner scintillators for CTR improvement introduces the degradation of detection efficiency.

Depth of interaction (DOI) correction has been proposed as another approach for reducing the effect of PTS without using thinner scintillators(Loignon-Houle *et al.* 2021). When scintillation photons are generated near the SiPM, two primary propagation paths to the SiPM exist: one is direct propagation to the SiPM, or the other is reflection from the far end of the scintillator to then reach the SiPM. Differences in these path lengths can be up to twice the scintillator length, which can result in a PTS on the order of tens of picoseconds. Therefore, identifying the interaction position and applying DOI correction can considerably improve the CTR. Detector configurations such as the side (Pourashraf *et al.* 2025; Cates and Levin 2018; Cates and Levin 2023; Pourashraf *et al.* 2021) and dual-ended readouts (Seifert and Schaart 2015; Pagano *et al.* 2024; Gan *et al.* 2025) have already been proposed as alternatives to conventional single-ended readout for reducing the PTS effects. Although these approaches improve CTR via DOI correction, several implementation challenges exist because of the doubled number of SiPMs used and the complex readout electronics, which can lead to higher costs. Therefore, a detector configuration is required to improve CTR while maintaining detection efficiency and minimizing cost.

To overcome these challenges, we previously proposed a new readout scheme named xDetector (Onishi and Ota 2025). The detector module of the xDetector included an LYSO crystal ($4 \times 4 \times 12.6$ mm$^3$) wrapped on four sides with an enhanced specular reflector (ESR) and coupled to a SiPM (S14161, Hamamatsu Photonics K.K., Japan) using optical grease, as illustrated in Figure 1(a). Three such detectors were aligned along the short axis to form a detector group (Figure 1(b)). Two such detector groups were orthogonally stacked with their bare sides facing each other via air coupling, and the outer surfaces were wrapped with Teflon tape to form the xDetector, as indicated in Figure 1(c). For convenience, the three channels are referred to as Ch1, Ch2, and Ch3 from left to right. A 3D-printed scintillator holder was used during the measurements to secure the xDetector configuration.

This configuration enabled the use of scintillators thinner than 20 mm to simultaneously reduce PTS and enhance LTE, thereby improving CTR while maintaining the equivalent detection efficiency compared to that of conventional PET detectors. Moreover, the xDetector enabled the development of high-performance TOF-DOI PET detectors with a smaller increase in the number of SiPMs than that used in the dual-ended readout. In our previous study, a CTR of $187.4 \pm 1.7$ ps FWHM was achieved using 4 mm pitch scintillators with an energy resolution of $\sim$10% FWHM. These results confirmed the potential of the xDetector to achieve both high detection efficiency and improved CTR. However, further improvement in the CTR through optimizations of detector components and timing correction based on interaction positions (typically referred to as DOI-based timing correction) is expected.

In this study, we explore the achievable CTR limit of the xDetector by employing an adequate choice of scintillators, photodetectors, and readout electronics, as well as through the application of DOI-based timing correction.

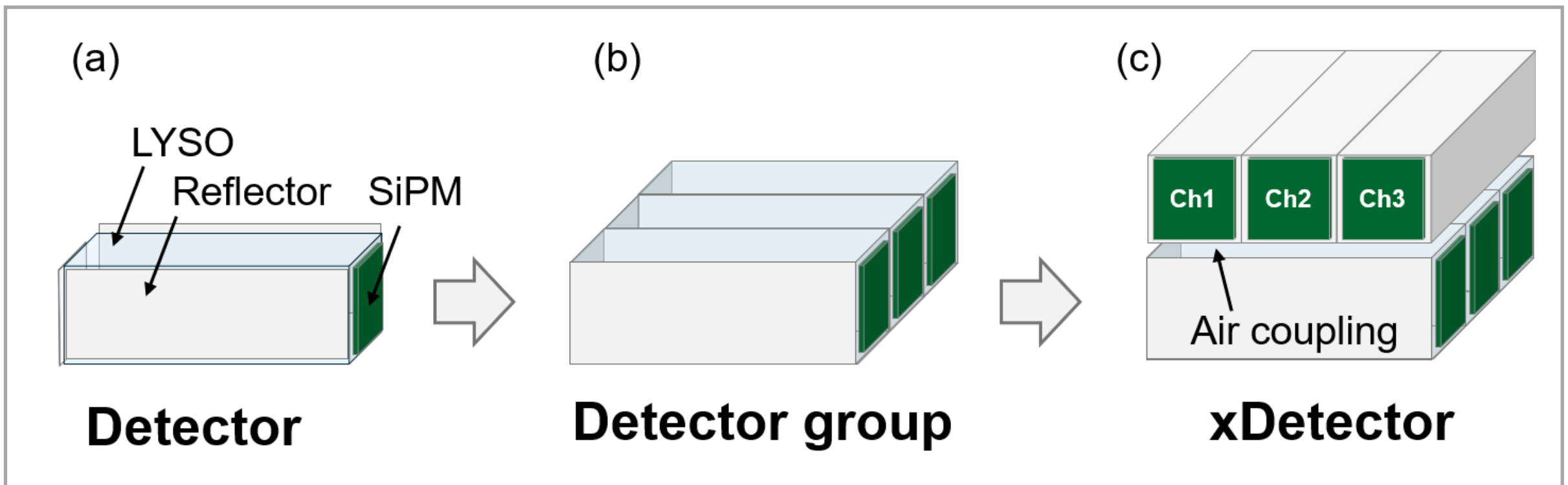

**Figure 1.** Schematic of the xDetector: (a) A detector consisting of LYSO crystals coupled to an SiPM using Meltmount, with the four sides covered by ESR. (b) A detector group formed by aligning three detectors along the short axis. (c) xDetector formed by orthogonally stacking two detector groups with their bare sides facing each other via air coupling.

## 2. Materials and methods

### 2.1. Detector optimization

As illustrated in Figure 2, three major improvements in detector components were implemented relative to that in the previous study (Onishi and Ota 2025) to improve CTR performance. Instead of the LSO crystals with a decay time of 46.6 ns used in the previous study, this study used faster LYSO crystals (CPI Inc., USA) with a decay time of 36.4 ns (Fig. 2(a)). Each decay time was measured using the time-correlated single-photon counting method.

Furthermore, conventional SiPMs (S14160, Hamamatsu Photonics, Japan) were replaced with new SiPMs (Hamamatsu Photonics, Japan)having an ~10% improved PDE at a wavelength of 420 nm. Spectroscopic PDE curves across the entire wavelength were plotted to measure the PDE improvement, as shown in Figure 2(b).

Although a high-frequency (HF) readout electronics was not used in the previous study, its use has been reported to improve CTR (Gundacker *et al.* 2020). Therefore, in this study, we designed a multichannel HF readout electronics based on two cascaded BGA616 amplifiers, as shown in Fig. 2(c).

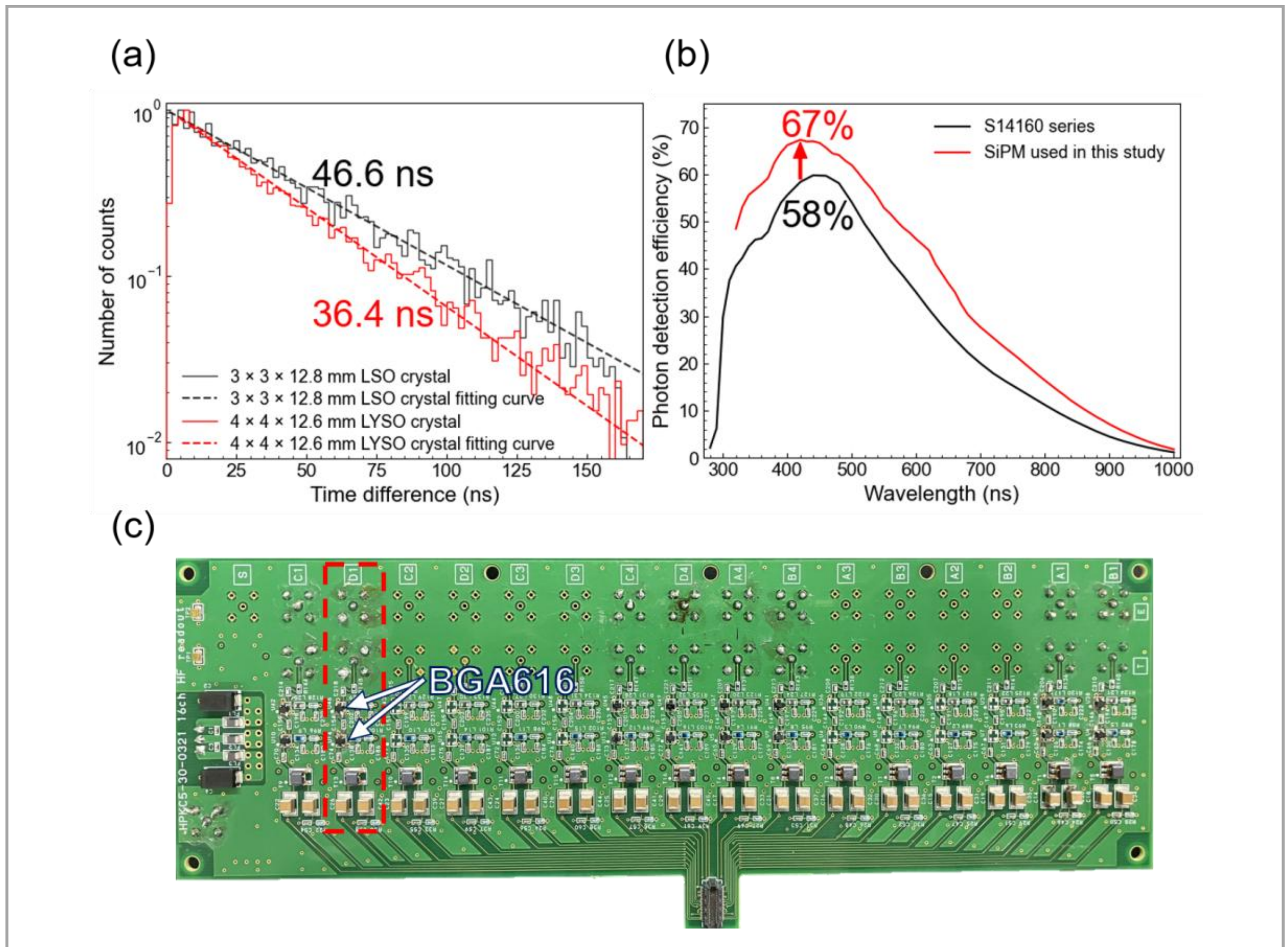

**Figure 2.** (a) The decay time of the LYSO crystal was 36.4 ns, which was an improvement of 10.2 ns compared to that in our previous study (46.6 ns) (Onishi and Ota 2025). (b) Spectral sensitivity curves of the SiPM used in this study and S14160 indicate an ~10% improvement in the PDE at a wavelength of 420 nm. (c) Photograph of the front side of the multichannel HF readout electronics based on two cascaded BGA616 amplifiers. The energy-signal readout electronics was implemented on the back side.

### 2.2. CTR measurement

The CTR was measured using the experimental setups displayed in Figure 3 for (a) the xDetector and (b) a conventional single detector to evaluate the performance of the xDetector. The reference detector included a LYSO crystal (4 × 4 × 8.4 mm$^3$, CPI Inc., USA) wrapped on five sides with ESR and Teflon tape, coupled to an SiPM of the same type as that used in the xDetector using Meltmount optical grease. The CTR of the reference detector was measured by placing two identical detectors face-to-face with a $^{22}$Na point source centered between them. The HF readout electronics was used for timing the signal readout. Consequently, the CTR of the reference detector was 101.8 ± 1.7 ps FWHM after optimizing the timing pick-off threshold level and bias voltage.

The xDetector was positioned facing the reference detector, with the $^{22}$Na point source placed at a sufficient distance to ensure uniform gamma-rays irradiation along the entire longitudinal side of the scintillator. For comparison with the xDetector, a conventional single detector was prepared comprising an LYSO crystal (4 × 4 × 12.6 or 20.0 mm$^3$, CPI Inc., USA) wrapped on five sides with ESR and Teflon tape, and it was coupled with the same SiPM as that used in the xDetector using Meltmount optical grease. The energy thresholds of all detectors were approximately set at the valley between the 511 keV photopeak and the Compton scattering region, assuming approximately 420 keV, based on the pulse height of the energy signal in the same manner as that in our previous study (Onishi and Ota 2025).

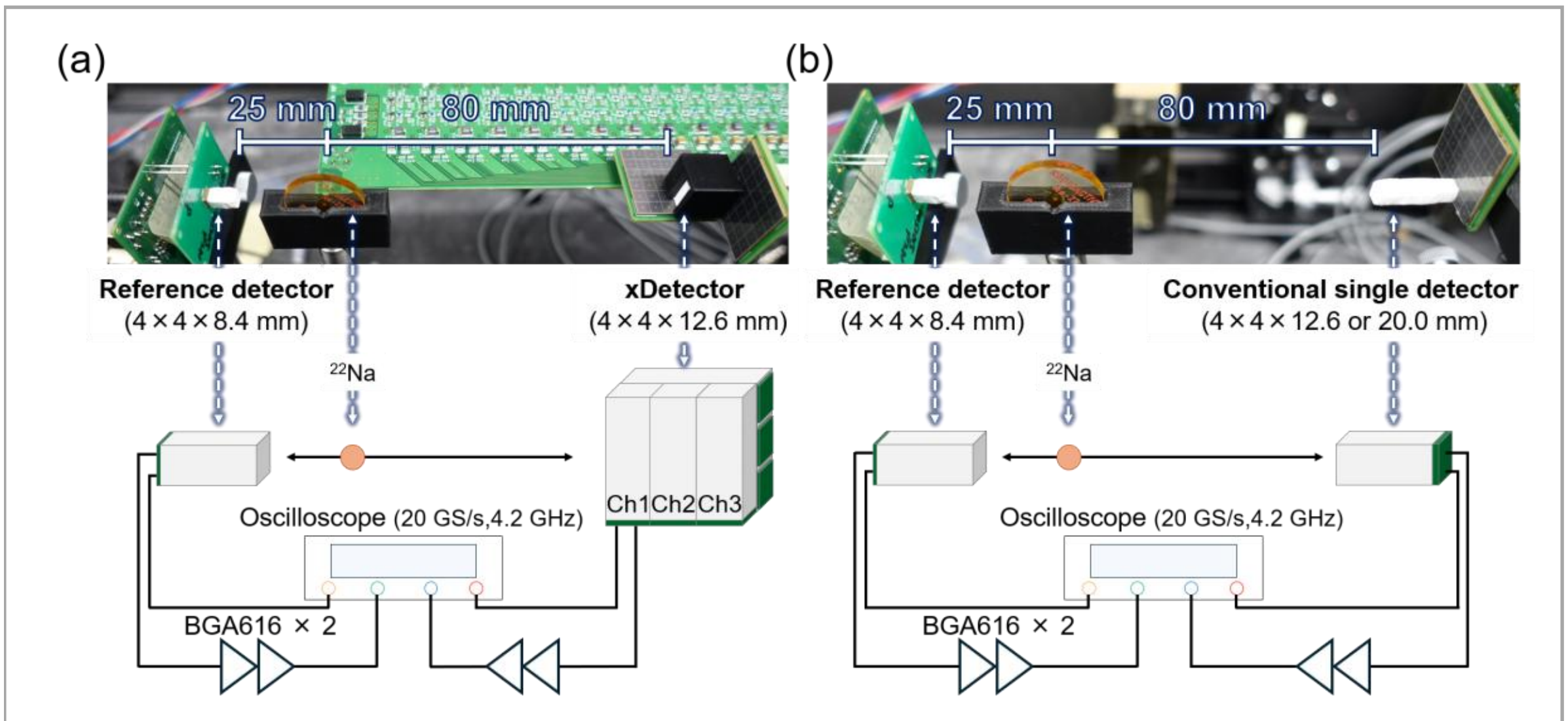

**Figure 3.** Experimental setups for the CTR measurements. (a) xDetector, (b) conventional single detector. The upper panels show photographs, and the lower panels show schematic diagrams.

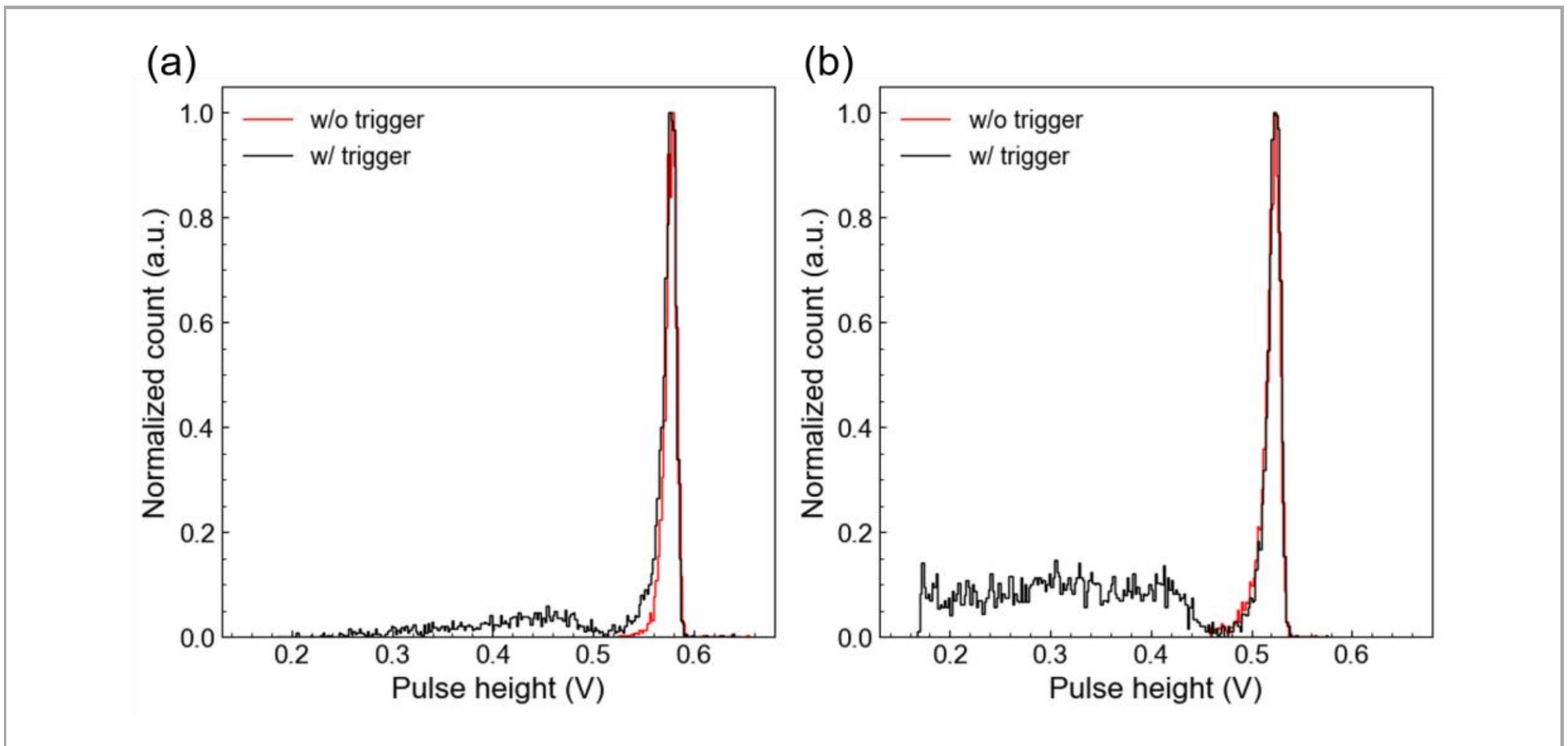

**Figure 4.** Energy histograms with and without trigger at a bias voltage of 64 V. (a) Reference detector and (b) Ch1 of xDetector. Energy thresholds were set at the valley between the 511 keV photopeak and Compton scattering region.

Figure 4 plots the energy histograms of (a) the reference detector and (b) Ch1 of the xDetector at the bias voltage of 64 V. Coincidence events were recorded using a digital oscilloscope (DSO-S 404A, Keysight, USA) operating at a sampling rate of 20 GS/s and a set bandwidth of 4.2 GHz. To maintain the maximum sampling rate, only timing signal channels were displayed, while two energy signals were used to trigger the oscilloscope. The CTR measurements were evaluated at bias voltages of 62, 63, and 64 V, corresponding over voltages from the breakdown voltage of 11, 12, and 13, respectively. For the analysis, baseline correction was applied to the timing signals of all coincidence events, and the threshold defining the detection time was optimized to obtain the best CTR. The time difference histograms between the reference detector and either the xDetector or the conventional single detector were fitted with a Gaussian function, and the FWHM of the obtained Gaussian was evaluated as the CTR.

### 2.3. DOI-based timing correction

The experimental setup for CTR measurement with DOI-based timing correction is shown in Figure 5. In this measurement, owing to the limited number of input channels of the oscilloscope, the interaction position was controlled by manually moving the xDetector instead of estimating it from the signals of channels aligned orthogonally to the longitudinal axis. The measurements were conducted using the Ch1 of the xDetector across four positions separated by 3.15 mm. For the sake of convenience, these positions were referred to as top, upper, lower, and bottom, ordered from the farthest to the nearest relative to the SiPM. The xDetector and reference detector were arranged face-to-face, with the $^{22}$Na point source positioned close to the xDetector to localize the gamma-ray irradiation region.

The energy thresholds were set to match those used in the CTR measurements at the same bias voltage. The detection timing threshold was set to an arbitrarily selected value under the same bias voltage. Moreover, time difference histograms corresponding to each position were shifted to an arbitrary reference value based on the mean value obtained from each Gaussian fitting and superimposed. The superimposed histogram was fitted with a Gaussian function, and the resulting FWHM was evaluated as the CTR with DOI-based timing correction.

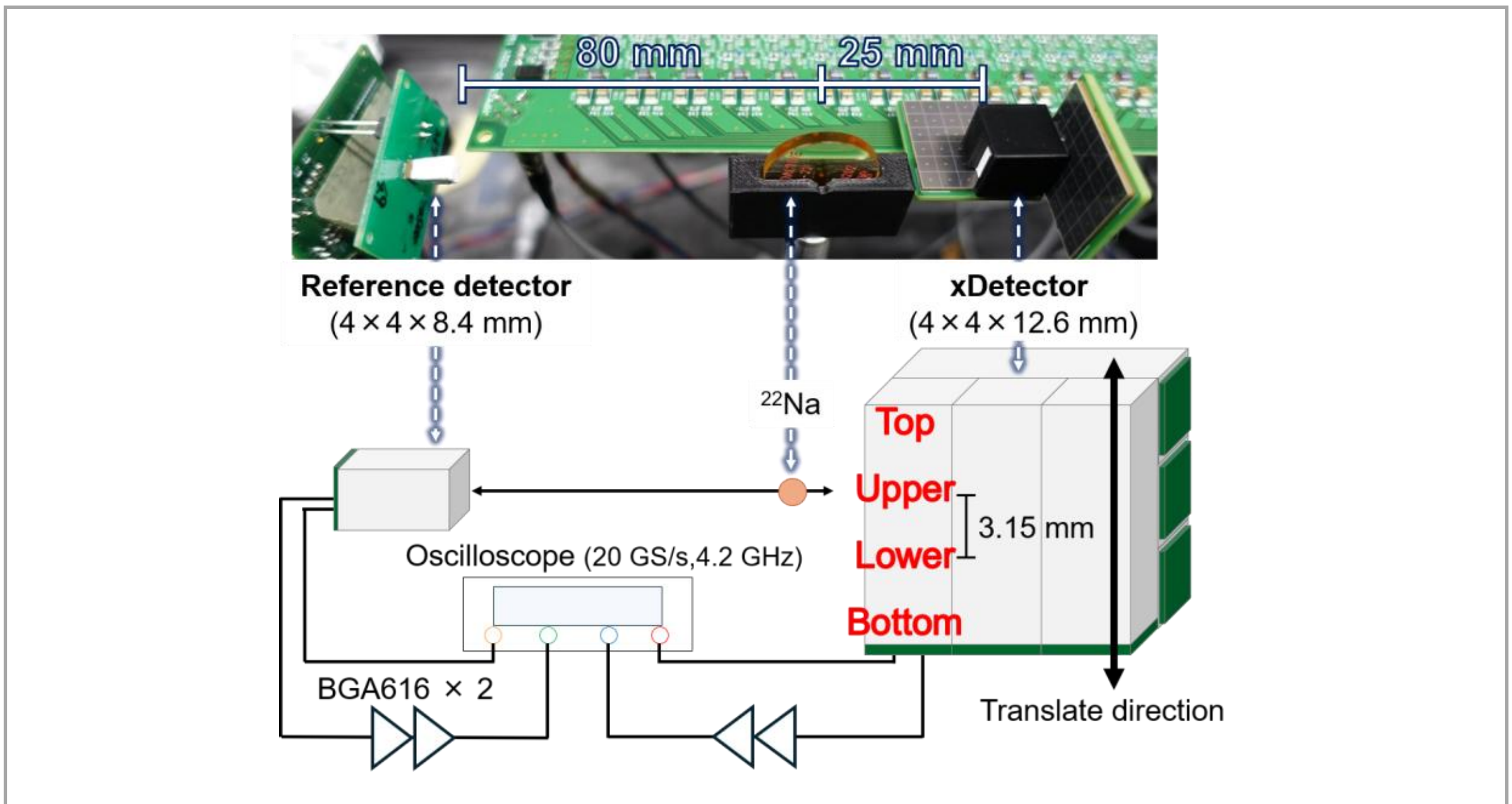

**Figure 5.** Experimental setup for the CTR measurement with DOI-based timing correction using the xDetector. The upper panels show photographs, and the lower panels show schematic diagrams.

# 3. Results

## 3.1 CTR measurement

The CTR values of the xDetector and conventional single detector measured under different bias voltages are shown in Figure 6. The CTR was measured between the xDetector or the conventional single detector and the reference detector. At a bias voltage of 63 V, the average CTR of the xDetector over all channels was 110.0 ps FWHM, which lay between those of the 12.6 and 20.0 mm scintillators of the conventional single detector. No significant differences in the CTR were observed among the individual channels of the xDetector. The best CTR of the xDetector was 107.8 $\pm$ 1.2 ps FWHM, recorded on Ch1 at the bias voltage of 63 V. Assuming a configuration with two xDetectors facing each other, the CTR was calculated to be 113.5 $\pm$ 2.7 ps FWHM.

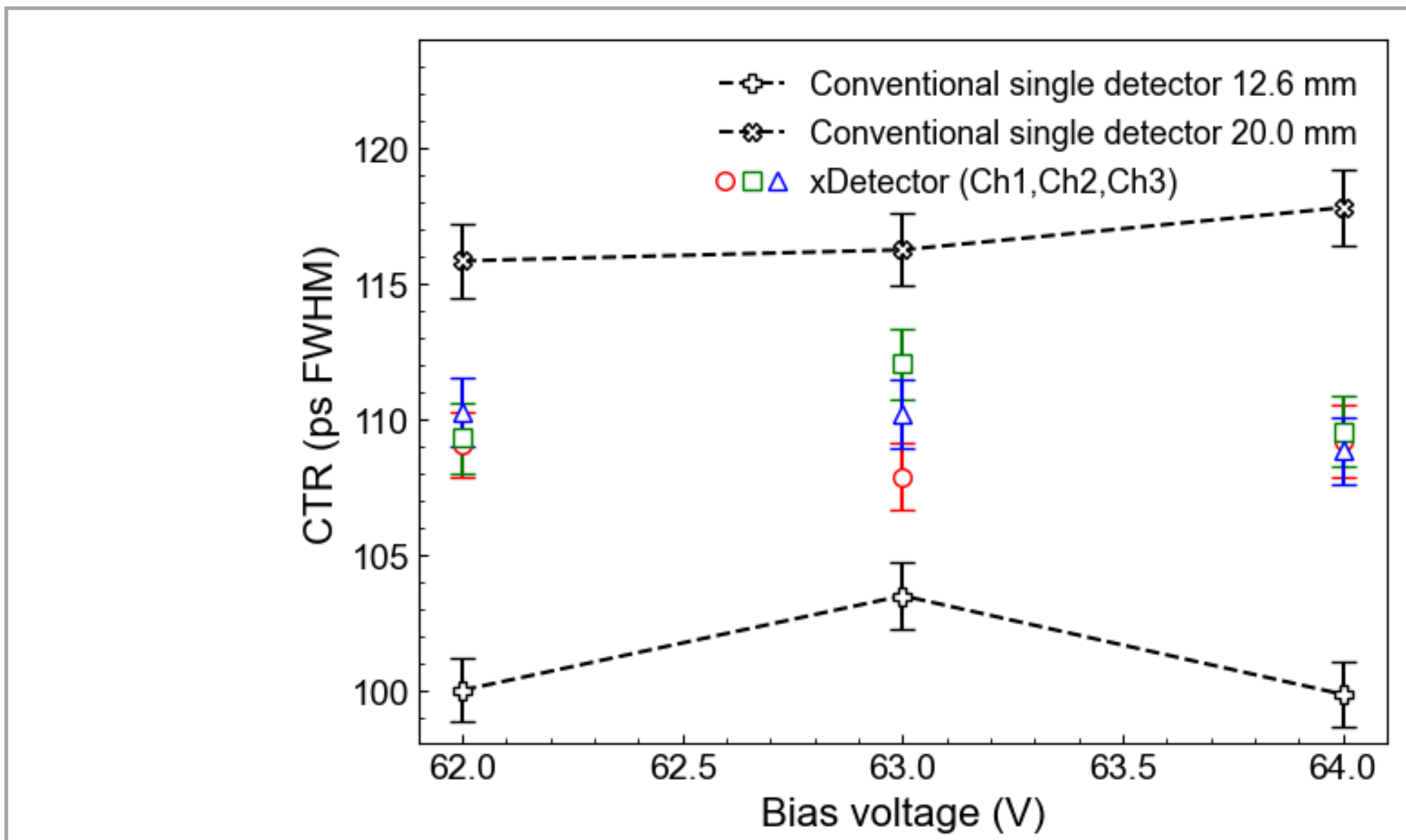

**Figure 6.** CTRs of the xDetector and conventional single detector at different bias voltages. The best CTR of the xDetector was 107.8$\pm$1.2 ps FWHM, which was measured in coincidence with the reference detector of 101.8$\pm$1.7 ps FWHM. Based on error propagation, the CTR for a configuration with two xDetectors facing each other was calculated to be 113.5$\pm$2.7 ps FWHM.

## 3.2 DOI-based timing correction

Figure 7 shows the CTR for each position under different bias voltages (a) and Gaussian fitting curves (b) for different positions at a bias voltage of 63 V. According to Figure 7(a), systematic differences in CTR were observed among positions; e.g., the top showed the fastest CTR of 99.6 $\pm$ 1.1 ps FWHM, whereas the bottom showed the slowest CTR of 110.5 $\pm$ 1.3 ps FWHM. Moreover, this trend is observed at other bias voltages. Figure 7(b) shows that the peak positions shift based on measurement positions ($T_{\text{xDetector}} - T_{\text{ref}}$); e.g., the top exhibited the slowest time difference, whereas the bottom showed the fastest value, with a slight separation between them. This trend was consistent regardless of the other bias voltages. The position dependent peak shift was attributed to optical propagation path lengths. DOI-based timing correction was applied to remove this effect.

As shown in Figure 8, the CTR of the conventional single detector and xDetector with and without DOI-based timing correction under different bias voltages. After applying the correction, the CTR was improved to ~105 ps FWHM. The trend of CTR as a function of bias voltage was similar independent of whether DOI-based timing correction was applied. The application of DOI-based timing correction resulted in an improvement of ~3 ps FWHM at all bias voltages. The best CTR achieved was 105.3 ± 0.6 ps FWHM at a bias voltage of 63 V. The CTR for a configuration with two xDetectors facing each other was 108.6 ± 1.9 ps FWHM. Moreover, compared with the conventional single detector using a 20.0 mm scintillator, CTR improved by an average of 10.3%, approaching that of the conventional single detector with the same thickness.

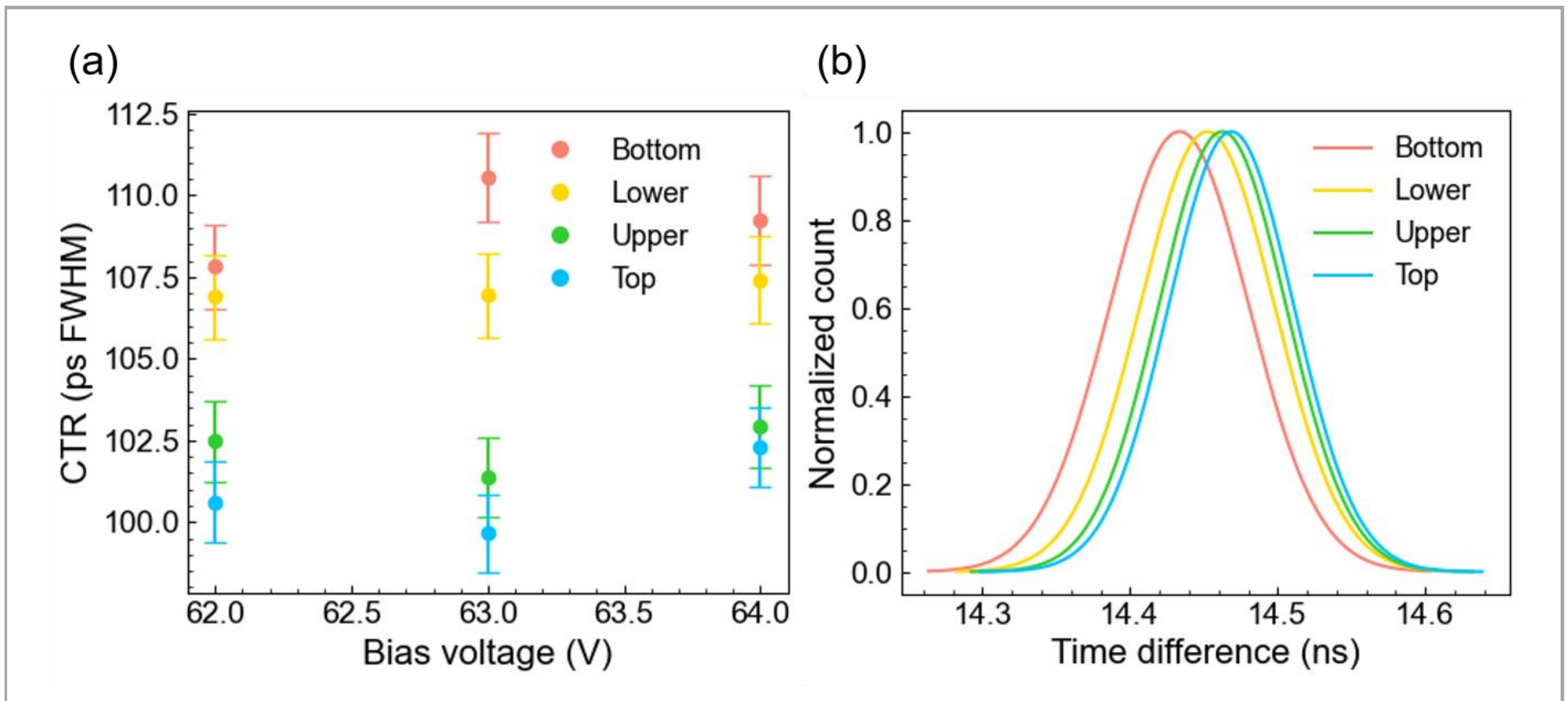

**Figure 7.** (a) CTR for each position under different bias voltages. (b) Fitting curves with a Gaussian function at a bias voltage of 63 V. These results were measured using the reference detector of 101.8 ± 1.7 ps FWHM.

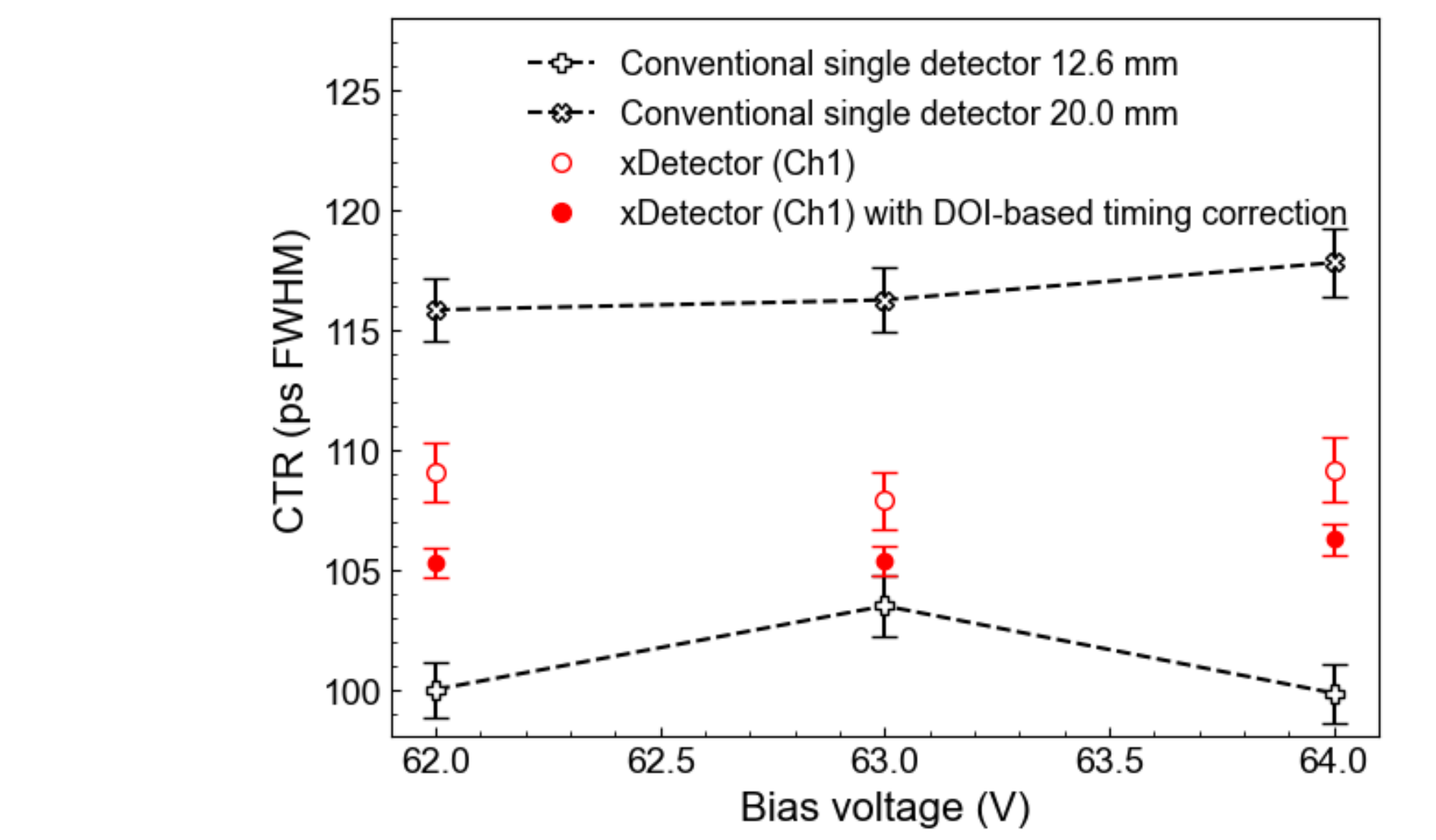

**Figure 8.** CTRs of the conventional single detector and xDetector with and without DOI-based timing correction under different bias voltages. It was measured while facing the reference detector, which obtained 101.8 ± 1.7 ps FWHM. The CTR was achieved 105.3 ± 0.6 ps FWHM at the bias voltage of 63 V by applying DOI-based timing correction, which corresponds to ~10% improvement compared to the conventional single detector with a 20.0 mm. Based on error propagation, the CTR for a configuration with two xDetectors facing each other was calculated to be 108.6±1.9 ps FWHM.

## 4. Discussion

In this study, we explored the potential CTR performance of the proposed xDetector by refining components such as the scintillators, SiPMs, and readout electronics, and by applying DOI-based timing correction. Consequently, the CTR of the two xDetectors in coincidence was 108.6 ± 1.9 ps FWHM, representing a substantial improvement compared with the 187 ps FWHM obtained in our previous proof-of-concept study (Onishi and Ota 2025).

As indicated in Figure 6, the CTR of the xDetector was better than that of the conventional single detector with a 20.0 mm scintillator. In contrast, the CTR of the xDetector was slightly inferior to that of the conventional single detector with a 12.6 mm scintillator; this can be attributed to the xDetector configuration, wherein one longitudinal side was not wrapped with ESR to retain spatial information by intentionally distributing scintillation photons. No significant differences in CTR were observed among the channels of the xDetector; however, the CTR of Ch2 was degraded by a few picoseconds FWHM. This slight degradation can be attributed to variations in optical coupling or the slight light leakage from Ch1 and Ch3.

The dependence of CTR on the measurement position within the scintillator is shown in Figure 7(b). The peak position of detection time becomes the fastest when the interaction position approaches the SiPM, and vice versa. This behavior can be ascribed to variations in the optical propagation path length within the scintillator, which introduces position-dependent variations in detection timing. Consequently, applying DOI-based timing correction at four positions improved the CTR by ~3 ps FWHM at each bias voltage; the best CTR achieved with the reference detector was 105.3 ± 0.6 ps FWHM, as shown in Figure 8. Based on error propagation, the CTR for a configuration with two xDetectors in coincidence was 108.6 ± 1.9 ps FWHM, which corresponded to an ~10% improvement compared to that in the conventional single detector with a 20.0 mm scintillator. These results confirm that further optimization of the detector geometry may bring CTR closer to the 100 ps FWHM range.

In contrast, a CTR of 102 ps FWHM has been reported using a side-readout scheme, wherein a 1 × 6 SiPM array was coupled to the longitudinal side of a 3 × 3 × 20 mm fast-LGSO (Mariscal-Castilla *et al.* 2024; Cates and Levin 2018; Cates and Levin 2023). Using this unique readout scheme, the CTR yielded an ~25% improvement from 137 ps FWHM to 102 ps FWHM, compared to the conventional single-ended-readout scheme. However, compared to the xDetector, the side-readout scheme requires approximately four times more SiPMs, which results in increased cost. In addition, the packing fraction defined as the ratio of the scintillator surface area facing the entire gamma ray was reduced in the side-readout scheme. These results confirm that the xDetector can improve CTR while maintaining cost efficiency, high packing fraction, and detection efficiency.

In this study, DOI-based timing correction was performed at four positions, corresponding to a longitudinal resolution of ~3 mm for the xDetector. However, in the previous study, the longitudinal resolution of the xDetector was reported to be ~6 mm. Considering this, the CTR was recalculated at two positions, and the corrected CTR was 106.4 ± 0.6 ps FWHM at a bias voltage of 63 V. This corresponds to a CTR of 110.7 ± 2.0 ps FWHM assuming two xDetectors facing each other. Although an improvement in CTR was observed even with two position corrections, improvement in the longitudinal resolution should be addressed to enhance CTR performance. Issues are maintaining the orthogonally stacked configuration unique to the xDetector and the slight increase in the number of SiPMs compared to the conventional PET modules. These issues require investigation because they may affect the packing fraction, cost, and assemble reproducibility.

## 5. Conclusion

In this study, we investigated the achievable CTR of the proposed xDetector. This detector concept was designed to overcome the trade-off between the CTR and detection efficiency by improving the scintillators, SiPMs, and the readout electronics, and by applying DOI-based timing correction. Consequently, the CTR was $107.8 \pm 1.2$ ps FWHM in coincidence with the reference detector. Applying the DOI-based timing correction improved the CTR further to 105.3 $\pm$ 0.6 ps FWHM. Based on error propagation, the CTR for the two xDetectors in coincidence was estimated to be $108.6 \pm 1.9$ ps FWHM, which corresponded to an $\sim$10% improvement compared to that of the conventional single detector with a 20.0 mm scintillator. The xDetector improved CTR while maintaining detection efficiency, thereby providing a promising design concept for PET detector modules. Future work can investigate the methods for enhancing longitudinal resolution, mechanically maintaining xDetector configuration, and reducing the number of SiPMs used towards further performance improvement.

## Acknowledgements

We are grateful to the members of the Solid State Division for providing SiPMs and to Mr. Akinori Saito for providing technical assistance.